\documentclass[prx,twocolumn,amsmath,amssymb,superscriptaddress,showpacs]{revtex4-2}

\usepackage[pdftex]{graphics}
\usepackage{graphicx}
\usepackage{color}
\usepackage[T1]{fontenc}
\usepackage{hyperref}

\begin{document}

\title{Giant natural optical rotation from chiral electromagnons in a collinear antiferromagnet}

\author{D. Maluski}
\author{M. Langenbach}
\affiliation{Institute of Physics II, University of Cologne, 50937 Cologne, Germany}
\author{D. Szaller}
\affiliation{HUN-REN-BME Condensed Matter Research Group, Budapest University of Technology and Economics, 
	1111 Budapest, Hungary}
\author{S. Reschke}
\author{L. Prodan}
\affiliation{Experimental Physics V, Center for Electronic Correlations and Magnetism, 
	University of Augsburg, 86159 Augsburg, Germany}
\author{I. C\'amara Mayorga}
\affiliation{Max Planck Institute for Radio Astronomy, 53121 Bonn, Germany}
\author{S.-W.~Cheong}
\affiliation{Rutgers Center for Emergent Materials and Department of Physics and Astronomy, 
	Rutgers University, Piscataway, New Jersey 08854, USA}
\author{V.~Tsurkan}
\affiliation{Experimental Physics V, Center for Electronic Correlations and Magnetism, 
	University of Augsburg, 86159 Augsburg, Germany}
\affiliation{Institute of Applied Physics, Moldova State University, MD-2028 Chisinau, 
	Republic of Moldova}
\author{I. K\'{e}zsm\'{a}rki}
\affiliation{Experimental Physics V, Center for Electronic Correlations and Magnetism, 
	University of Augsburg, 86159 Augsburg, Germany}
\author{J. Hemberger}
\author{M. Gr\"{u}ninger}
\affiliation{Institute of Physics II, University of Cologne, 50937 Cologne, Germany}

\date{December 5, 2023}

\begin{abstract}
In NiTe$_3$O$_6$ with a chiral crystal structure, we report on a giant natural 
optical rotation of the lowest-energy magnon. This polarization rotation, as large 
as 140$^\circ$/mm, corresponds to a path difference between right and left circular 
polarizations that is comparable to the sample thickness. 
Natural optical rotation, being a measure of structural chirality, is highly unusual 
for long-wavelength magnons. The collinear antiferromagnetic order of  NiTe$_3$O$_6$ 
makes this giant effect even more peculiar: Chirality of the crystal structure does 
not affect the magnetic ground state but is strongly manifested in the lowest excited 
state. We show that the dynamic magnetoelectric effect, turning this magnon to a 
magnetic- and electric-dipole active hybrid mode, generates the giant natural optical 
rotation. In finite magnetic fields, it also leads to a strong optical magnetochiral 
effect. 
\end{abstract}

\maketitle

Chirality, as first termed by Kelvin, is a structural or geometrical property. 
A chiral object cannot be overlayed with its mirror image, it has only pure rotation 
symmetries and cannot possess an inversion center or mirror planes.  
A widely used tool for detecting chirality is natural optical activity, i.e., 
a rotation of linear polarization of light. It originates from the different velocities 
of right and left circularly polarized (\textit{rcp/lcp}) photon polaritons in chiral media, 
i.e., circular birefringence.  Via Kramers-Kronig relations, this is tied to 
circular dichroism, a difference in absorption for \textit{rcp} and \textit{lcp} light.

Chirality emerges in diverse material platforms. 
It is a key property of living matter that also governs its functionalities, 
as prominently manifested in the homo-chirality of life \cite{Sallembien22,Sasselov23,Ozturk23}, 
with only left-handed amino acids existing in nature.
In inorganic crystals, chirality is the source for a range of fascinating phenomena 
\cite{Chang18,Fecher22,Felser23,Rikken97,Bordacs12,Kezsmarki14,Zhu18,Nomura19,Yokouchi20,Guo22}.
We restrict our discussion to the interplay with magnetism \cite{Cheong22}. 
Chirality of the crystal structure can induce chirality of magnetic textures, 
leading to, e.g., spin helices and skyrmions \cite{Muehlbauer09,Tokura21}. 
Such chiral spin textures may also emerge spontaneously in achiral crystals due to 
competing exchange interactions \cite{Kurumaji19,Hirschberger19}. 
Remarkably, chirality can arise from the superposition of achiral lattice symmetry 
and achiral magnetic order \cite{Bordacs12,Saito08}. This provides the possibility 
for magnetic control of chirality in materials hosting simple magnetic orders.

Recently, a plethora of non-reciprocal effects has been reported in chiral systems 
that are magnetic or exposed to a magnetic field \cite{Tokura18,Cheong18}. 
Magnetochiral dichroism refers to the difference in light absorption between 
counter-propagating beams with wavevector $\mathbf{q}$ along and 
opposite to the magnetization or magnetic field and has been reported for magnon, 
phonon, and electronic excitations 
\cite{Rikken97,Zhu18,Bordacs12,Train08,Atzori20,Atzori21,Szaller13,Kezsmarki11,Takahashi12,Takahashi13,Kezsmarki14,Kuzmenko15,Kezsmarki15,Yu18,Sirenko21,Yokosuk20,Park22,Sessoli15}. 
Magnetochiral anisotropy has also been observed in dc electric and thermal 
transport \cite{Pop14}, most recently in the mysterious kagome compound CsV$_3$Sb$_5$ \cite{Guo22,Wu22}. 
Electromagnetic induction of a helix is another prominent example for functionalities 
governed by chirality \cite{Yokouchi20}.

The term chirality (handedness) has also been extended to dynamical processes such 
as quasiparticle excitations 
\cite{Onoda08,Jenni22,Daniels18,Nambu20,Liu22,Sahasrabudhe23,Ishito23}. In the case 
of magnons, handedness refers to right and left circular polarizations. 
In a ferromagnet, magnons are typically right circular polarized, corresponding 
to clockwise precession of the magnetization \cite{Onoda08,Jenni22}.
In contrast, in a collinear easy-axis antiferromagnet like Ni$_3$TeO$_6$, 
right- and left-handed magnons are degenerate in zero field such that a 
superposition of the two degenerate modes yields eigenstates that are 
not well characterized by handedness. 
The chiral crystal structure preserves this degeneracy. 
Concerning the optical response of the degenerate magnons to \textit{rcp} 
and \textit{lcp} light, we note that \textit{the} typical source of natural 
optical rotation, the non-local effect of spatial dispersion \cite{Glazer06}, 
is relevant in the visible frequency range but 
can be neglected for magnons with energies in the GHz-THz regime where the wavelength 
of light is several orders of magnitude larger. 
However, as we show in the following, the $\Gamma$-point magnons still may cause giant 
natural optical rotation in a chiral crystal due to the dynamic magnetoelectric effect.

Here, we report on the giant natural optical rotation of the lowest-energy magnon 
in the collinear antiferromagnet Ni$_3$TeO$_6$ with chiral crystal structure (R3). 
The stunning strength of the effect is well illustrated by the optical path 
difference between \textit{rcp} and \textit{lcp} light being similar to the 
sample thickness. 
We show that this magnon mode picks up the chirality of the lattice via the dynamic 
magnetoelectric effect, leading to the observed natural optical rotation.
In the long-wavelength limit, applicable to our THz data, the light-matter interaction 
is restricted to uniform excitations of electric and magnetic dipole moments. 
Correspondingly, spatial dispersion can be neglected and the only way to generate 
sizable natural optical rotation is the interference between electric and magnetic 
dipole excitations, the so-called dynamic magnetoelectric effect.
Despite the hybrid nature, the electromagnon in Ni$_3$TeO$_6$ has an extraordinarily 
long lifetime in the nanosecond range, which implies an unusual mechanism of the 
magnetoelectric coupling. In finite magnetic field, where the collinear 
antiferromagnetic order is still preserved, this magnon also exhibits magneto-optical 
rotation and magnetochiral birefringence. 
In contrast to the former, the latter shares a common origin with the natural 
optical rotation, i.e., it is governed by the same magnetoelectric coefficient 
and is getting activated by the Zeeman splitting of the doubly-degenerate 
magnon resonance.

\begin{figure}[tb]
	\centering
	\includegraphics[width=\columnwidth]{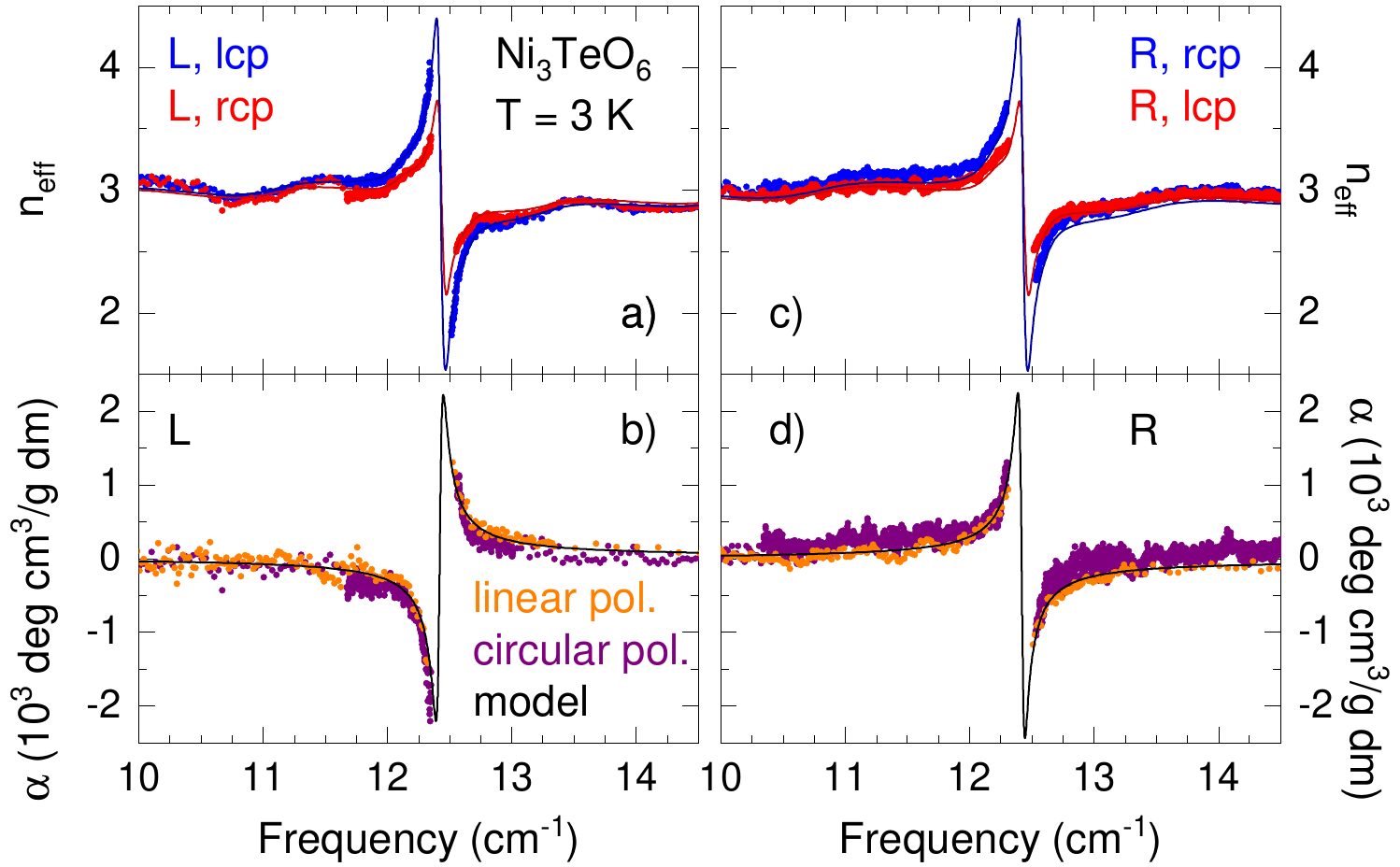}
	\caption{\textbf{Giant natural optical rotation of Ni$_3$TeO$_6$ at $T$\,=\,3\,K.\@}
 	a),c) Effective refractive index $n_{\rm eff}$\,=\,$[L_{\rm opt}(\omega)/d]+1$ 
 	measured with \textit{rcp} and \textit{lcp} light on samples with opposite helicity, 
 	L and R, where $L_{\rm opt}$ denotes the measured optical path difference and 
 	$d$ the sample thickness. 
    Curves with the same color are related by symmetry, e.g., 
 	$n_{rcp}^{\rm R}$\,=\,$n_{lcp}^{\rm L}$.
	b),d) Specific polarization rotation $\alpha$ measured with linear polarization 
	(orange) and as determined from 
	$\alpha_{\rm eff}$\,=\,$(\pi/\lambda\rho)(n_{{\rm eff},rcp}-n_{{\rm eff},lcp})$ 
	measured with circular polarization (purple) on a sample with given helicity, 
	cf.\ a).   
	Solid black lines are based on Eq.\ (\ref{eq:Dn}) for 
	$\omega_0$\,=\,12.42\,cm$^{-1}$, $|\omega_{xx}|$\,=\,0.0375\,cm$^{-1}$, 
	and $\gamma$\,=\,0.05\,cm$^{-1}$. 
	}
	\label{fig:n}
\end{figure}

\textbf{Natural optical activity in zero field --}
We study the terahertz response of insulating Ni$_3$TeO$_6$ in transmission geometry 
for light propagation along the chiral $c$ axis in the collinear antiferromagnetic phase 
at $T$\,=\,3\,K, well below the ordering temperature 
$T_N$\,=\,52-55\,K \cite{Zivkovic10,Oh14,Yokosuk16}.
In chiral Ni$_3$TeO$_6$, the eigenstates for light are right and left circular polarized 
(\textit{rcp} and \textit{lcp}), and we compare the response for circular polarization 
with the optical rotation for linear polarization. 
Figure \ref{fig:n} depicts results for both sample helicities, R and L, at 3\,K in zero magnetic field. 
The excitation with eigenfrequency $\omega_0$\,=\,12.42\,cm$^{-1}$ (or 372.3\,GHz, 1.540\,meV) 
has been interpreted as an electromagnon since broad features were observed at similar frequencies 
in Raman results and time-domain terahertz data of ceramic samples \cite{Skiadopoulou17}. 
Our data reveal the giant natural optical activity of this mode, underlining its chiral character.  
In Fig.\ \ref{fig:n}, panels b) and d) show the specific rotation $\alpha$\,=\,$\theta/\rho d$ 
for linear polarization (orange), where $\theta$ denotes the measured polarization rotation, 
$\rho$\,=\,6.4\,g/cm$^3$ is the density \cite{Becker06}, and $d$ is the sample thickness. 
As expected, the sign of $\alpha$ is inverted between the two sample helicities.
With the actual thicknesses being $d_{\rm R}$\,=\,0.78\,mm and $d_{\rm L}$\,=\,1.05\,mm, 
strong absorption suppresses the amplitude below the noise level in a region of about 
$\pm$\,0.1\,cm$^{-1}$ around $\omega_0$, i.e., we are missing data at the very maximum of absorption. 
The measured $\alpha$ is extraordinarily large, exceeding 1000$^\circ$\,cm$^3$/g\,dm 
at $\omega_0 \pm$\,0.1\,cm$^{-1}$.
The corresponding fit (see below) peaks at $2.2\cdot 10^3$\,deg\,cm$^3$/g\,dm. 
For $d$\,=\,1\,mm, this corresponds to $\theta$\,$\approx$\,140$^\circ$, 
indeed a giant rotation for a collinear antiferromagnet in zero magnetic field.

We focus on the frequency range close to the resonance, where circular birefringence is 
accompanied by circular dichroism. Consequently, linearly polarized light does not only 
experience a rotation of the polarization plane but becomes elliptically polarized or 
even circularly polarized. In the latter limit, the polarization rotation is an 
ill-defined quantity. Therefore, we confirm our result using circular polarization.
The rotation of linear polarization $\theta$\,=\,$(\pi d/\lambda)(n_{rcp}-n_{lcp})$
is caused by circular birefringence, i.e., a difference of the real parts of the refractive 
indices, where $\lambda$ denotes the vacuum wavelength. Using circular polarization, 
our measured THz phase data probe $n_{rcp}$ and $n_{lcp}$ individually. 
Panels a) and c) of Fig.\ \ref{fig:n} plot the effective refractive index $n_{\rm eff}(\omega)$ 
derived from the measured optical path difference
$L_{\rm opt}(\omega)$\,=\,$[n_{\rm eff}(\omega)\!-\!1]\,d$ 
induced by the sample in comparison to vacuum. 
Neglecting multiple reflections within the sample, $n_{\rm eff}(\omega)$ equals $n(\omega)$ 
for a given circular polarization. 
For frequencies not too close to the absorption feature, such multiple reflections 
cause the small Fabry-P\'erot interference fringes in $n_{\rm eff}$ 
with a period of $\Delta \omega$\,=\,$(2nd)^{-1} \approx\,$1.6\,cm$^{-1}$ (2.1\,cm$^{-1}$) 
in the L (R) sample. Panels b) and d) compare the specific polarization rotation $\alpha$ 
measured with linear (orange) and circular polarization (purple), the latter determined via 
$\alpha_{\rm eff}$\,=\,$(\pi / \lambda \rho)(n_{{\rm eff},rcp}-n_{{\rm eff},lcp})$ 
for a given sample helicity from the data plotted in a) and c). 
Note that the interference fringes nearly drop out by taking the difference. 
The two results for linear and circular polarization agree very well with each other, 
both for the L and the R domain.

\begin{figure}[tb]
	\centering
	\includegraphics[width=\columnwidth]{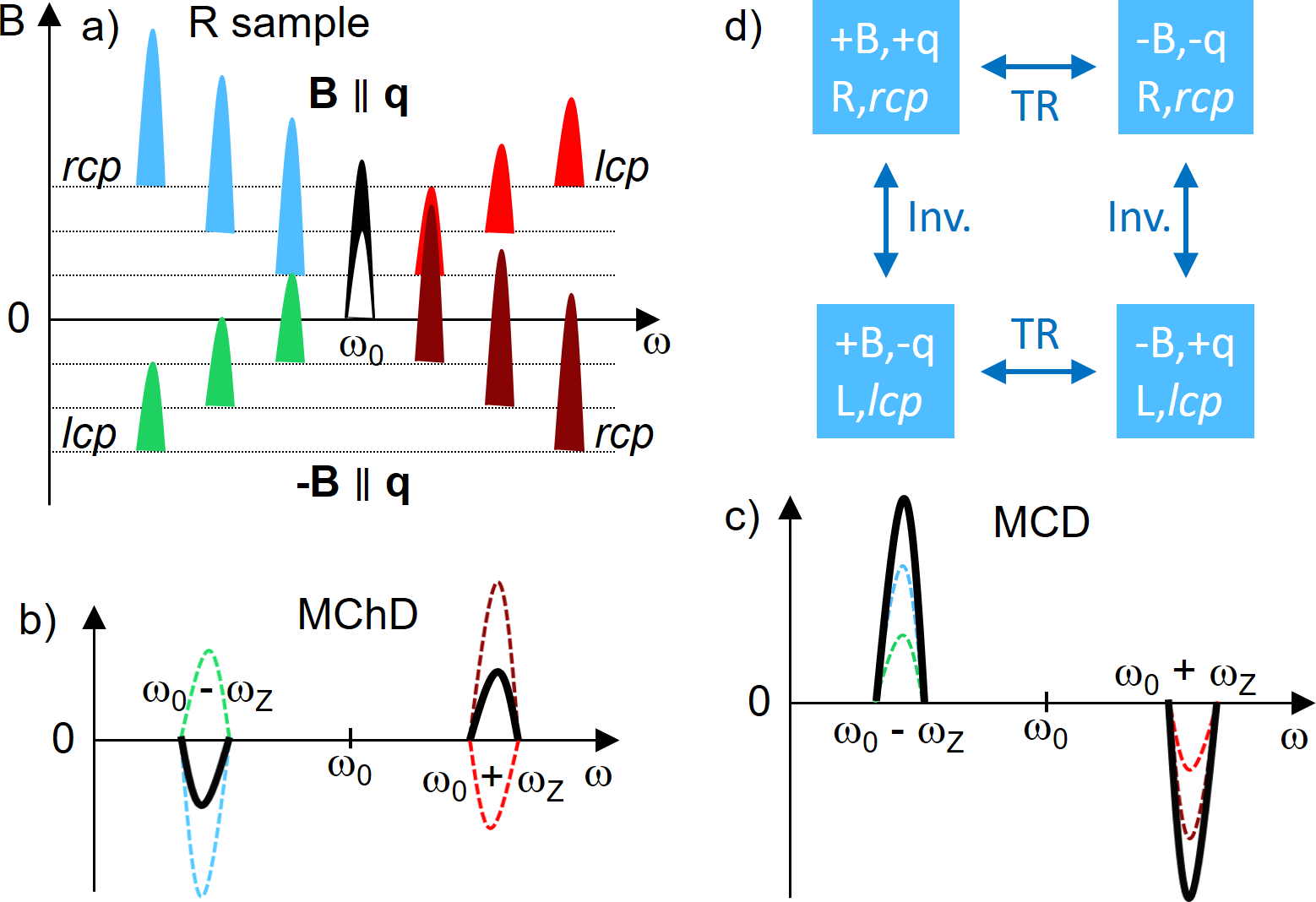}
	\caption{\textbf{Sketch of behavior in magnetic field.} 
	For simplicity, we depict simple peaks, and their intensity difference denotes dichroism, 
	which via Kramers-Kronig relations is equivalent to birefringence.	
	a) In zero field, different peak intensities $I_{rcp}$ and $I_{lcp}$ at $\omega_0$ 
	(black, white) depict natural circular dichroism. Finite magnetic field (anti-)parallel 
	to the chiral axis yields a Zeeman splitting, which directly results in magnetochiral 
	dichroism (MChD) and magnetocircular dichroism (MCD) sketched in b) and c), 
	cf.\ Eqs.\ (\ref{eq:MCB}) and (\ref{eq:MChB}). 
	The four colors reflect quadrochroism equivalent to MChD, cf.\ Fig.\ \ref{fig:1T}. 
	b),c) MCD and MChD (thick solid lines) measure the sum and difference of the 
	peaks at $\omega_0-\omega_Z$ compared to the sum and difference at $\omega_0+\omega_Z$. 
	Both MCD and MChD vanish in zero field for $\omega_Z$\,=\,0. 
	d) Applying time reversal (TR) and/or spatial inversion (Inv.) yields a 
	family of four equivalent settings with identical responses, 
	shown here for the blue feature in a).  		
	}
	\label{fig:sketch_B}
\end{figure}

\begin{figure}[tb]
	\centering
	\includegraphics[width=\columnwidth]{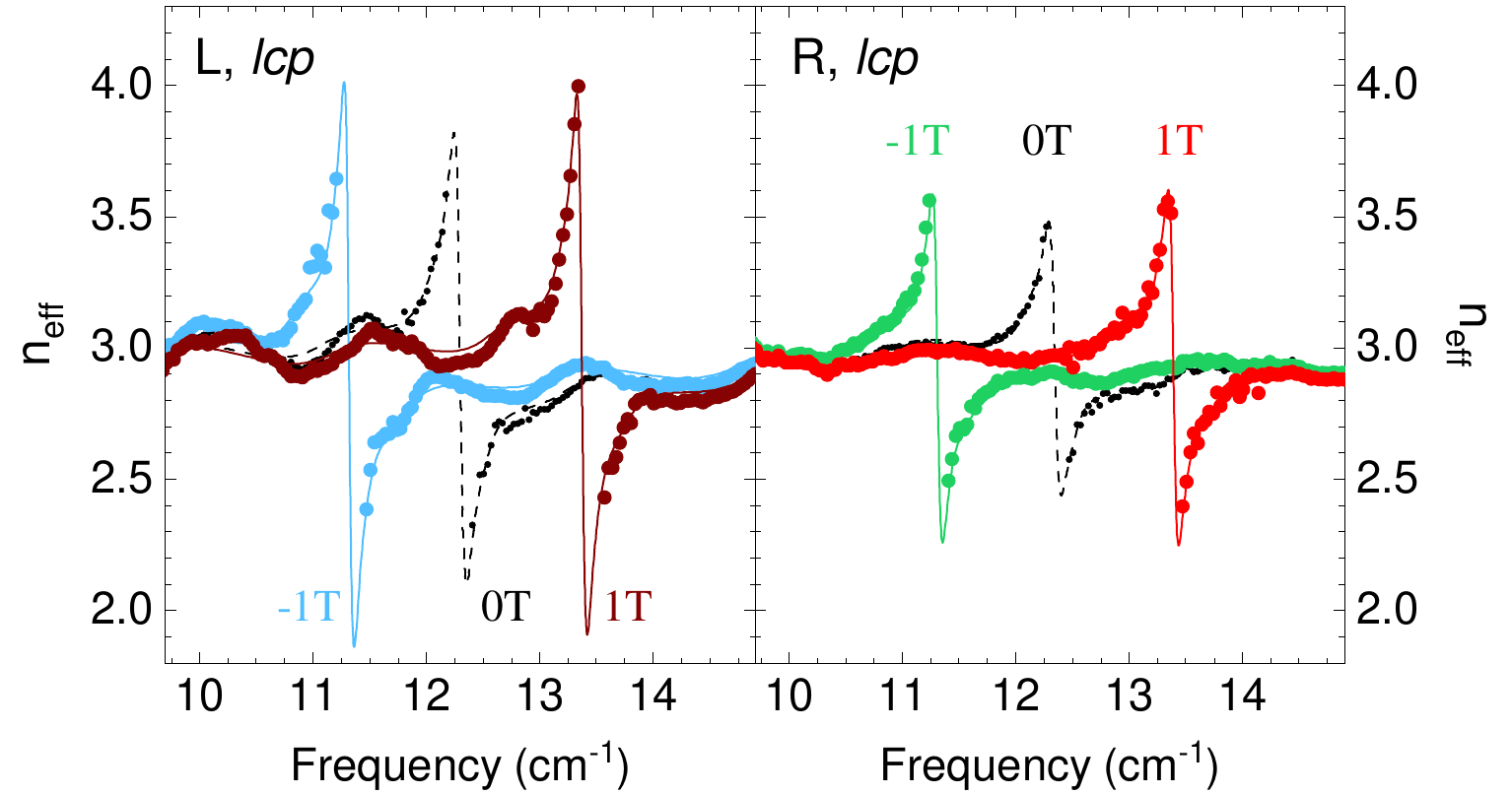}
	\caption{\textbf{Zeeman splitting of the electromagnon in Ni$_3$TeO$_6$ for $|B|$\,=\,1\,T.} 
		All data sets are measured at 13.5\,K with \textit{lcp} light and the same direction 
		of $\mathbf{q}$, parallel to the $c$ axis. 
		The combination of natural circular birefringence and Zeeman splitting yields magnetochiral 
		birefringence equivalent to quadrochroism. The four different responses in our data arise 
		based on the direction of $\mathbf{B}$, where $\pm$1\,T denotes 
		$\pm \mathbf{B} \parallel \mathbf{q}$, 
		and the two different sample helicities, L and R (left and right panels). 
		The four responses can be classified by the eigenfrequency 
		$\omega_0 \pm \omega_Z$ and the oscillator strength in the refractive index. 
		The color code follows Fig.\ \ref{fig:sketch_B}a). 
	}
	\label{fig:1T}
\end{figure}

Now we address the quantitative analysis. 
In Ni$_3$TeO$_6$, any sizable chiral response in the terahertz range has to arise from 
the dynamical magnetoelectric coupling tensor $\hat{\chi}^{me}$, as stated above. 
For light propagation along the trigonal $c$ axis, we hence may consider both the 
dielectric function $\varepsilon(\omega)$ and the permeability $\mu(\omega)$ as 
scalar quantities, see Sect.\ \ref{sec:oscimodel}. 
Solving Maxwell's equations for the point group $31^\prime$ of the 
collinear antiferromagnetic phase and $\mathbf{q} \parallel c$, 
we find the complex refractive index 
\begin{equation}
	\label{eq:solution_N}
	N_{rcp/lcp}(\omega) = 
	\sqrt{\varepsilon\, \mu +  \left( \chi^{me}_{xy} \right)^2 } 
    \mp i \, \chi^{me}_{xx}  
\end{equation} 
with $N$\,=\,$n+i\,\kappa$. 
This expression quantifies how the chiral character, detected by the natural optical rotation 
$\alpha\! \propto \! \Re(N_{rcp} - N_{lcp})$, arises based on the 
magnetoelectric tensor, i.e., for a finite imaginary part of the diagonal component $\chi^{me}_{xx}$ 
that connects $B_x(\omega)$ and $E_x(\omega)$.
Considering a single mode and assuming a Lorentzian shape, time-reversal symmetry implies 
\begin{eqnarray}
    \chi^{me}_{ij}(\omega) & = & \frac{i \, \omega \, \omega_{ij}}{\omega_0^2 - \omega^2 - i \gamma \omega }  
\end{eqnarray}
where $\omega_{xx}$ and $\omega_{xy}$ govern the strength of the magnetoelectric response 
and $\gamma$ denotes the damping. Note that this expression for $\chi^{me}_{ij}$ 
guarantees that the \textit{static} linear magnetoelectric response vanishes, 
as required. For the real-valued difference $n_{rcp}-n_{lcp}$ we find
\begin{equation}
	\label{eq:Dn}
	\Re(N_{rcp} - N_{lcp}) = 
	\frac{2\, \omega \, \omega_{xx}\, (\omega_0^2 - \omega^2)}{(\omega_0^2 - \omega^2)^2 + (\gamma \omega)^2} \, .
\end{equation}
A fit of $\alpha$ measured at 3\,K yields $\omega_0$\,=\,12.42\,cm$^{-1}$ and 
$\omega_{xx}$\,=\,-0.039\,cm$^{-1}$ (0.036\,cm$^{-1}$) for the L (R) sample, 
see Figs.\ \ref{fig:n}b), d). 
These results for $\omega_0$ and $\omega_{xx}$ are not affected by the uncertainty 
of $\gamma$, for which we estimate $\gamma$\,=\,0.05\,cm$^{-1}$ 
(or 1.5\,GHz, 6\,$\mu$eV) with $\gamma/\omega_0$\,$\approx$\,0.004. 
We cannot exclude a still slightly smaller value since we are missing data at 
the absorption maximum in the range $\omega_0 \pm$\,0.1\,cm$^{-1}$. 
This yields a rough estimate of the lifetime in the nanosecond range, a stunning 
result. 
The line widths of electromagnons in other compounds are much broader, 
with typical values of $\gamma/\omega_0$\,$\gtrsim$\,0.1 in, e.g., 
Li$M$PO$_4$ ($M$\,=\,Fe, Co, Ni) or 
TbFe$_3$(BO$_3$)$_4$ \cite{Peedu22,Peedu19,Kocsis18,Szaller17}. 
Still larger values have been observed in, e.g., $R$MnO$_3$, also in inelastic 
neutron scattering \cite{Pimenov06,Valdes09,Finger14}, 
see \cite{Tokura14} for a review.  
A large width agrees with the intuition for a hybrid mode that contains 
both magnetic and electric-dipole-active contributions, offering different 
decay channels.
Comparably small values of $\gamma$\,$\approx$\,1\,cm$^{-1}$ and  
$\gamma/\omega_0$\,$\approx$\,0.02-0.03 were reported for Co$_2$Mo$_3$O$_8$ 
and Fe$_2$Mo$_3$O$_8$ \cite{Reschke22,Csizi20}. 
For electromagnons, our result $\gamma/\omega_0 \approx 0.004$ is outstanding. 
The small line width observed in Ni$_3$TeO$_6$ rather is comparable to results 
for pure magnon modes studied in collinear antiferromagnets with the 
neutron resonance spin echo technique \cite{Bayrakci13}. 
In fact, the line width is only an order of magnitude larger than that of magnons in the 
reference low-loss spin-wave material yttrium iron garnet as observed in optical transmission 
experiments \cite{Weymann21}.

The very small linewidth $\gamma$ also drives the very large specific polarization 
rotation $\alpha$. The fit using Eq.\ (\ref{eq:Dn}) allows us to determine the maximum value 
at about {$\omega_0 \pm \gamma/2$, where $\alpha$ reaches 2200$^\circ$\,cm$^3$/g\,dm at 3\,K.\@ 
This corresponds to a maximum difference 
$|n_{rcp}-n_{lcp}|_{\rm max}$\,$\approx$ $\omega_{xx}/\gamma$\,$\approx$\,3/4 such that 
the relative optical path delay $\delta L_{\rm opt}$\,=\,$(n_{rcp}-n_{lcp})\,d$ between 
the two polarizations nearly equals the sample thickness, again highlighting the giant 
strength of the effect. Also in the visible frequency range, a large value of 
$\alpha_{\rm VIS}$\,=\,$1355^\circ$\,cm$^3$/g\,dm has been reported \cite{Wang15}.  
We emphasize that the corresponding value of 
$n_{rcp}-n_{lcp} \propto \alpha \, \lambda$ 
is orders of magnitude smaller in the visible than in the terahertz range.

In passing we mention that $n_{rcp}$ may turn negative, as discussed 
for chiral metamaterials \cite{Zhang09}. 
In the present case, this would be achieved for $\gamma \leq$\,0.012\,cm$^{-1}$ 
or for an increase of $\omega_{xx}$ by a factor 5. 
Another noteworthy aspect is the asymptotic behavior of $\theta$.  
For large $\omega \! \gg \! \omega_0$, the difference $|n_{rcp}-n_{lcp}|$ extrapolates to 
$2\omega_{xx}/\omega$. 
This yields the counterintuitive result of a constant \textit{high}-frequency extrapolation 
$\theta(\omega\!=\!\infty)$\,=\,$2\pi d \,\omega_{xx}$ with $d$ in units of cm and 
$\omega_{xx}$ in cm$^{-1}$. For $d$\,=\,1\,mm, our model predicts 
$|\theta(\infty)|$\,$\approx$\,1.4$^\circ$. 
In fact, this has to be compensated by other excitations such that the total high-frequency 
limit of the rotation $\theta$ vanishes \cite{Smith76}.

\textbf{Nonreciprocal directional birefringence in finite magnetic field --}
In the collinear antiferromagnetic phase at 3\,K in zero magnetic field, considering a given 
sample helicity, there are two degenerate electromagnons at $\omega_0$\,=\,12.42\,cm$^{-1}$.  
Due to $\chi_{xx}^{me}$, they show opposite helicities such that one 
contributes to $n_{rcp}$, the other one to $n_{lcp}$, 
together causing the giant natural optical activity. 
Application of an external magnetic field $\mathbf{B}$  along the chiral axis 
breaks time-reversal symmetry and induces a Zeeman splitting between these 
two modes, as shown in Fig.\ \ref{fig:1T} for $B$\,=\,$\pm 1$\,T at 13.5\,K 
and sketched in Fig.\ \ref{fig:sketch_B}a). The Zeeman splitting of the eigenfrequencies 
$\omega_0 \pm \omega_Z$ with  
$\omega_Z$\,=\,$g\,\mu_B |B|/(\hbar 2\pi c)$ in units of cm$^{-1}$ reveals 
the $g$ factor $g$\,$\approx$\,2.2. 
A value of 2.26 has been found in an ESR study on a polycrystalline sample \cite{Zupan71}.
Furthermore, the Zeeman splitting causes magnetocircular birefringence 
(MCB), i.e., Faraday rotation, see Fig.\ \ref{fig:sketch_B}c), 
\begin{equation}
	\label{eq:MCB}
	n_{\rm MCB}   =  \left(n_{rcp}^{+q} + n_{lcp}^{-q}\right)  
	               - \left(n_{rcp}^{-q} + n_{lcp}^{+q}\right)  
\end{equation}
where $+q$ and $-q$ refer to $\mathbf{q}$ parallel and antiparallel 
to $\mathbf{B}$. 
The result shown in Fig.\ \ref{fig:MCB} and the sketch in Fig.\ \ref{fig:sketch_B}c) 
illustrate that MCB measures the difference between the responses at $\omega_0 -\omega_Z$ 
and $\omega_0 + \omega_Z$. 
Note that MCB changes upon time reversal but is unaffected upon spatial inversion, 
i.e., it appears in magnetic field also in the absence of natural circular birefringence. 
In contrast, magnetochiral birefringence (MChB, cf.\ Figs.\ \ref{fig:sketch_B}b) and \ref{fig:MCB})  
\begin{eqnarray}
	\label{eq:MChB}
 n_{\rm MChB} & =  & \left(n_{rcp}^{+q} - n_{lcp}^{-q}\right)  - \left(n_{rcp}^{-q} - n_{lcp}^{+q}\right) 
 \\
 \label{eq:NDD}
 & = & \left(n_{rcp}^{+q} - n_{rcp}^{-q} \right)  + \left(n_{lcp}^{+q} - n_{lcp}^{-q} \right)
\end{eqnarray}
is inverted upon either time reversal or spatial inversion. It arises in our data due 
to the Zeeman splitting of chiral modes showing circular birefringence, 
as emphasized by Eq.\ (\ref{eq:MChB}) and sketched in Fig.\ \ref{fig:sketch_B}b). 
Note that for such electromagnon excitations, natural optical birefringence and circular dichroism 
are two sides of the same coin in terms of Kramers-Kronig relations connecting real and imaginary 
parts of a linear response function. MChB is equivalent to non-reciprocal directional birefringence, 
i.e., different properties for counterpropagating electromagnetic waves as highlighted 
by Eq.\ (\ref{eq:NDD}).

\begin{figure}[tb]
	\centering
	\includegraphics[width=\columnwidth]{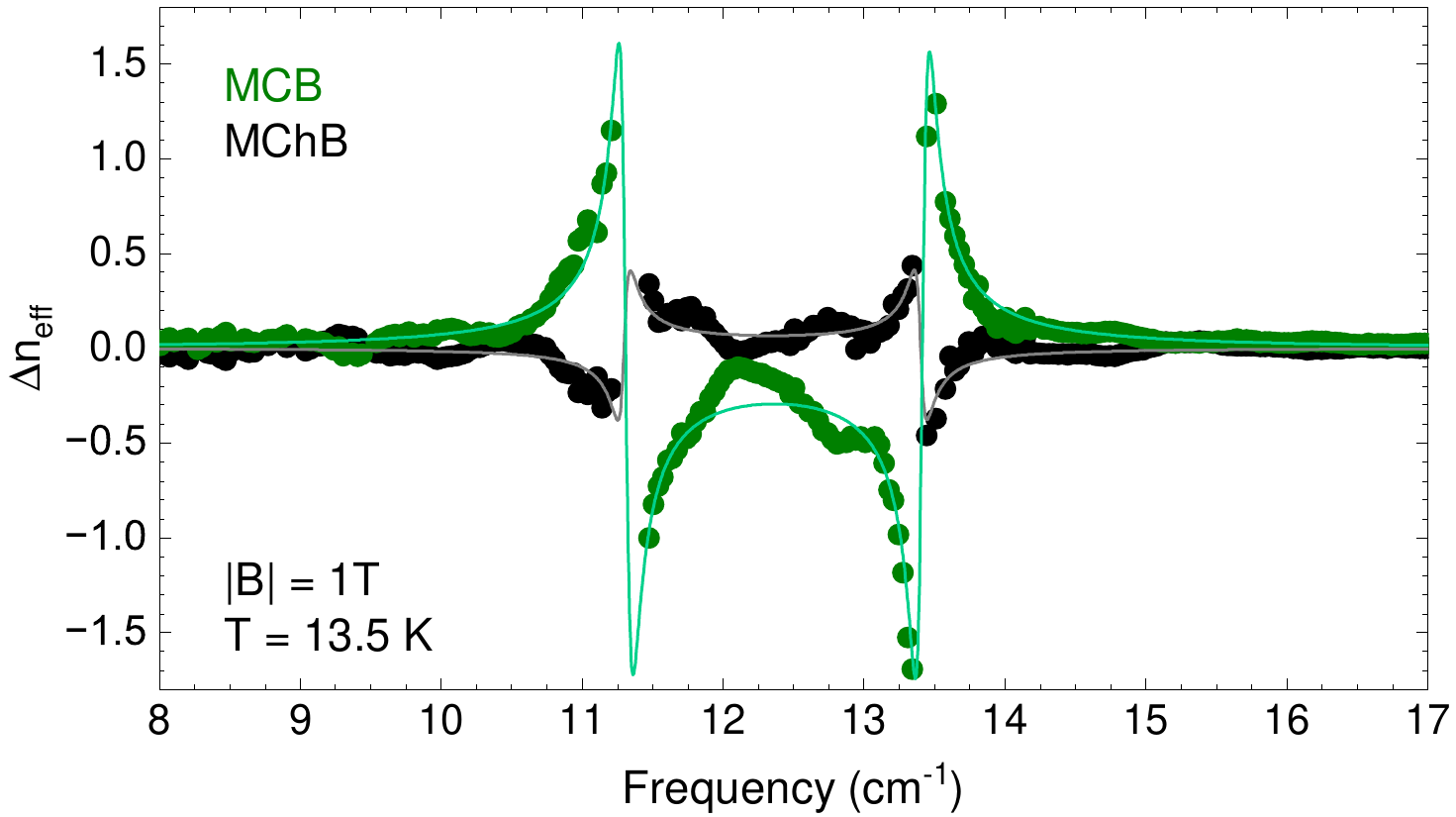}
	\caption{\textbf{Magnetocircular birefringence (MCB) and magnetochiral birefringence 
	(MChB) for $|B|$\,=\,1\,T at 13.5\,K.} 
    The data (symbols) were derived from the results shown in Fig.\ \ref{fig:1T} 
    using Eqs.\ (\ref{eq:MCB}), (\ref{eq:MChB}), and (\ref{eq:lcpq}). 
    The Fabry-P\'erot interference fringes nearly cancel out for frequencies below 
    and above the two features, but they cause some deviation from the expected line 
    shape in between the two peaks. 
    Solid lines depict a fit based on Eq.\ (\ref{eq:solution_N}) plus Zeeman splitting, 
    neglecting the fringes. 
    }
	\label{fig:MCB}
\end{figure}

Non-reciprocal directional dichroism is also called quadrochroism, referring to 
the four different values of $N_{rcp/lcp}^{\pm q}$ for a given sample helicity. 
Concerning magnons, it typically occur in complex magnetic structures 
that are affected by the external magnetic field 
\cite{Szaller13,Kezsmarki11,Takahashi12,Bordacs12,Takahashi13,Kezsmarki14,Kuzmenko15,Kezsmarki15}.
In contrast, we consider the intuitive case of a collinear antiferromagnet with chiral 
crystal structure, in which the existence of quadrochroism is straightforward. 
Starting from natural circular dichroism, the simple role of the magnetic field is to lift 
the degeneracy of the chiral magnons, hence quadrochroism appears based on the Zeeman effect.
More explicitly, we consider circular polarization, sample helicity, and the orientations 
of $\mathbf{B}$ and $\mathbf{q}$ along the chiral axis. This yields 16 possible variants 
$N_{rcp/lcp,{\rm L/R}}^{\pm q,\pm B}$. Applying time reversal and/or spatial inversion to the 
entire experimental setup yields four families, each hosting four members with 
equivalent properties.  One example is given in Fig.\ \ref{fig:sketch_B}d). 
Quadrochroism means that these four families exhibit different properties. 
The four quantities $\Re(N_{rcp/lcp}^{\pm q})$ in Eq.\ (\ref{eq:MChB}) refer to a fixed 
orientation of $\mathbf{B}$ and fixed sample helicity and represent the four different families.

For measurements in finite $B$, our setup is restricted to \textit{lcp} light. 
Observation of the four different responses hence requires to study both domains, L and R.\@ 
The data in Fig.\ \ref{fig:1T} depict the four different cases 
$n_{lcp,{\rm L/R}}^{+ q,\pm B}$ that are equivalent to $n_{rcp/lcp,{\rm L}}^{\pm q,+B}$ 
according to 
\begin{eqnarray}
	n_{\rm MChB} & =  & n_{rcp,{\rm L}}^{+q,+B} - n_{lcp,{\rm L}}^{-q,+B}  
	                  - n_{rcp,{\rm L}}^{-q,+B} + n_{lcp,{\rm L}}^{+q,+B} 
	\\
\label{eq:lcpq}
	& = & n_{lcp,{\rm R}}^{+q,-B} - n_{lcp,{\rm L}}^{+q,-B}  
	    - n_{lcp,{\rm R}}^{+q,+B} + n_{lcp,{\rm L}}^{+q,+B}  \, ,
\end{eqnarray}
as schematically depicted in Fig.\ \ref{fig:sketch_B}b). Similarly, n$_{\rm MCB}$ can also 
be determined from measurements on both L and R domains with \textit{lcp} light.  
The corresponding spectra of n$_{\rm MChB}$ and n$_{\rm MCB}$ are plotted in Fig.\ \ref{fig:MCB}.

Finite $B$ breaks time-reversal symmetry and allows for additional terms in the tensors 
$\hat{\varepsilon}$, $\hat{\mu}$, and $\hat{\chi}^{me}$ such as an off-diagonal 
contribution to $\hat{\varepsilon}$ reflecting the Faraday effect. 
We cannot separate the effects of all additional parameters and hence refrain from a 
quantitative microscopic analysis in finite field. 
For simplicity, we stick to Eq.\ (\ref{eq:solution_N}) where we add the Zeeman splitting, 
resulting in the solid lines plotted in Fig.\ \ref{fig:MCB}. 
For the parameters at 13.5\,K we find $\omega_0$\,=\,12.35\,cm$^{-1}$, 
$\omega_Z$\,=\,1.05\,cm$^{-1}$, 
$|\omega_{xx}|$\,=\,0.035\,cm$^{-1}$, and $\gamma$\,=\,0.09\,cm$^{-1}$. 
Compared to 3\,K, the values of $\omega_0$ and $\omega_{xx}$ are only slightly reduced 
at 13.5\,K, while the linewidth is enhanced by almost a factor of 2.

In conclusion, the dynamic magnetoelectric tensor is the source of the giant natural 
optical rotation in the collinear antiferromagnet Ni$_3$TeO$_6$ with chiral crystal structure. 
In a straightforward way, it also yields a strong magnetochiral effect in finite magnetic field. 
The mechanism strongly differs from the typical case based on spatial dispersion. 
It also differs from previous reports of natural optical activity of magnons in other compounds 
where a large effect arises in a large magnetic field \cite{Bordacs12,Kuzmenko14,Kuzmenko19,Iguchi21}.

The stunning strength of 140$^\circ$/mm rotation for linear polarization is 
based on the very narrow line width of 1.5\,GHz. This value is much smaller 
than typically reported for electromagnons in other compounds. 
The experimental determination of such a narrow line width hinges on the fact that 
the frequency resolution of our continuous-wave THz spectrometer is by far superior 
to the values that are typically achieved with the commonly used time-domain setups. 
The combination of excellent frequency resolution, broadband use of circular 
polarization, and phase-sensitive detection sets a benchmark for optical 
spectroscopy of (electro-) magnons.

\section{Materials and Methods}
\label{sec:exp}

\subsection{Samples}
\label{subsec:samples}

The compound Ni$_3$TeO$_6$ crystallizes in a trigonal corundum-related structure with  
space group $R3$ \cite{Becker06,Sankar13,Wang15}. 
It is chiral and polar even at room temperature, a rare combination. 
Chirality arises due to ordered distortions of the NiO$_6$ octahedra, 
while the polarization is due to a displacement of Ni and Te ions from the center 
of the surrounding oxygen octahedra \cite{Wang15}.  
The compound exhibits two enantiomers which correspond to growth twins or domains 
that can be transformed into each other by inversion. 
Growth and characterization of the L sample has been described in 
Refs.\ \cite{Oh14,Wang15}. It shows both enantiomers, 
and we covered the R pieces using a polarization microscope. 
The single-crystalline R sample of Ni$_3$TeO$_6$ has been grown using the 
chemical transport reaction method. As initial material for the crystal growth, 
we used polycrystalline Ni$_3$TeO$_6$ prepared by solid state reactions 
from binary NiO and TeO$_2$. Anhydrous TeCl$_4$ was utilized as a transport agent. 
The growth was performed in a two-zone furnace in a temperature gradient 
from 740 to 680$^\circ$C.

With the nominal valence Ni$^{2+}$, this Mott-insulating material hosts spin 
$S$\,=\,1 on three inequivalent Ni sites. Collinear antiferromagnetic order (AFM-I) 
with the propagation vector $(0,0,1/2)$ sets in at 
$T_N$\,=\,52-55\,K \cite{Zivkovic10,Oh14,Yokosuk16}. 
The ordered moments form ferromagnetic planes that are stacked 
antiferromagnetically along the $c$ axis \cite{Zivkovic10,Oh14}. 
At 5\,K, a spin-flop transition to the AFM-II phase has been reported for 
B\,=\,8.83\,T applied along the $c$ axis \cite{Oh14,Yokosuk16,Kim15}.
Usually, the combination of antiferromagnetic order and electric polarization is discussed 
in terms of multiferroicity. Strictly speaking, however, Ni$_3$TeO$_6$ does not belong to 
the class of multiferroics since the structure is polar already in the paramagnetic state. 
The electric polarization is not induced by magnetic order but it is strongly enhanced 
in the magnetically ordered state due to magnetoelectric coupling \cite{Oh14}.

Concerning the symmetry in the AFM-I phase, we may stick to the ordinary magnetic space groups 
due to the collinear character of magnetic order. 
Starting from $R3$, only $R_{R}3$ (No.\ 1242 or 146.3.1242) describes antiferromagnetic order, 
it hence is the correct magnetic space group for the AFM-I phase. 
The corresponding point group is $31'$. It contains time reversal $1'$ since the combination 
of $1'$ and $\frac{1}{2}\mathbf{c}$ translation remains a symmetry operation of the space group 
in the antiferromagnetic state.

\subsection{Continuous-wave terahertz measurements}
\label{subsec:thz}

The essential experimental aspects of our study are the usage of circularly polarized light 
with high frequency resolution and phase-sensitive detection. This is achieved with a 
continuous-wave spectrometer based on photomixing 
\cite{Roggenbuck10,Roggenbuck12,Roggenbuck13,Maluski22}. 
The beams of two tunable near-infrared lasers ({\sc Toptica}) are superimposed in a 
fiber array and illuminate photomixers -- proprietary technology from the 
Max Planck Institute for Radio Astronomy.
Using phase-sensitive homodyne detection and two fiber stretchers for 
phase modulation, we measure both amplitude and phase of the terahertz radiation. 
Comparing the results on the sample with those of a reference measurement on an empty aperture, 
the phase delay $\varphi(\omega)$\,=\,$\varphi_{\rm sam}(\omega)-\varphi_{\rm ref}(\omega)$ 
corresponds to the optical path difference 
$L_{\rm opt}(\omega)$\,=\,$\varphi(\omega) \lambda/2\pi$\,=\,$(n_{\rm eff}(\omega)\!-\!1)\,d$  
induced by the sample in comparison to vacuum. 
The photomixers emit and detect left circularly polarized 
(\textit{lcp}) light \cite{Camara06,Furuya09,Garufo16}. 
In transmission geometry, data for right circular polarization (\textit{rcp}) were collected 
using a linear polarizer in front of the sample and a stainless steel plate behind the sample, 
reflecting the light under 45$^\circ$ incidence to transform the transmitted \textit{rcp} 
component to \textit{lcp}. 
Alternatively, the polarization rotation was determined using a linear polarizer in front 
of the sample and a second one operating as analyzer behind the sample. The analyzer was 
rotated in steps of $5^\circ$ in the vicinity of the magnon and in steps of $10^\circ$ 
otherwise. 
The setup achieves a frequency resolution in the MHz range. In the vicinity of the 
electromagnon, data were collected with a typical step size of 
$\Delta \omega$\,=\,50\,MHz (0.0017\,cm$^{-1}$), otherwise $\Delta \omega$\,=\,500\,MHz was used. 
In zero magnetic field, data were measured at 3\,K in an optical $^4$He bath cryostat on 
an optical table with four lenses forming two Gaussian telescopes, focusing the beam 
onto the sample \cite{Maluski22}. 
For measurements in magnetic field up to 8\,T, 
we employ fiber-coupled photomixers that we have engineered 
for cryogenic operation, allowing us to operate the photomixers inside a windowless 
magneto-cryostat. Using face-to-face geometry with a distance of about 6\,cm between 
the photomixers, the sample is located in the focus of the photomixers' 
Si lens that couples the terahertz radiation to free space. 
Inside the magnet, reference and sample can be exchanged using a piezo-rotator. 
Data in magnetic field were collected at 13.5\,K where the setup is most stable, 
which is related to the heat carried into the magnet by the two laser beams.

The magnon eigenfrequency $\omega_0$\,=\,12.42\,cm$^{-1}$ corresponds to a vacuum wavelength 
$\lambda_{\rm vac} \approx 0.8$\,mm. This is comparable to the sample size. 
Therefore, we focus on the phase data which is less sensitive to possible diffraction effects 
for measurements on samples in a frequency range where the sample size is not much larger 
than the vacuum wavelength $\lambda_{\rm vac}$ \cite{Maluski22}. 
Roughly speaking, diffraction reduces the detected amplitude but the photons that arrive 
at the detector still carry meaningful phase information. 
Similar to the phase data, the measurement of the rotation of linear polarization is 
not sensitive to a diffraction-induced suppression of the absolute value of the terahertz amplitude.

\subsection{Oscillator model}
\label{sec:oscimodel}

Assuming a Lorentzian shape of the excitation, time-reversal symmetry implies 
\begin{eqnarray}
	\varepsilon(\omega) = \varepsilon_\infty & +  & \frac{\omega_p^2}{\omega_0^2 - \omega^2 - i \gamma \omega }  
	\\
	\mu(\omega) = 1 & + & \frac{\omega_m^2}{\omega_0^2 - \omega^2 - i \gamma \omega }  
	\\
	\chi^{me}_{ij}(\omega) & = & \frac{i \, \omega \, \omega_{ij}}{\omega_0^2 - \omega^2 - i \gamma \omega }  
	\\	\label{eq:constraint}
	{\rm with} \,\,\,\, & & 
	\omega_p^2 \omega_m^2 / \omega_0^2 = \omega_{xx}^2 + \omega_{xy}^2 \, ,
\end{eqnarray}
where $\omega_0$ denotes the eigenfrequency, $\gamma$ is the linewidth, and 
$\omega_{xx}$ and $\omega_{xy}$ govern the strength of the magnetoelectric response 
in analogy to the plasma frequency $\omega_p$ and the equivalent parameter $\omega_m$ for 
the magnetic dipole response, while $\varepsilon_\infty$ summarizes the contributions 
from higher energies. With the typical choice $\mu_\infty$\,=\,1, such high-energy 
contributions are neglected in $\mu(\omega)$. 
\\

\begin{acknowledgments}
We gratefully acknowledge fruitful discussions with P. Becker and L. Bohat\'{y}  
as well as funding from the Deutsche Forschungsgemeinschaft (DFG, German Research Foundation) 
via Project numbers 277146847 (CRC 1238, project B02) 
and 492547816 (TRR 360). 
SWC was supported by the W. M. Keck foundation grant to the 
Keck Center for Quantum Magnetism at Rutgers University. 
DS acknowledges the support of the Janos Bolyai Research Scholarship (BO-00580/22/11) 
and that of the the New National Excellence Program (ÚNKP-23-5-BME-414).
VT acknowledges support via the project ANCD 20.80009.5007.19 (Moldova). 
\end{acknowledgments}


\begin{thebibliography}{99}
	
\bibitem{Sallembien22}
Q. Sallembien, L. Bouteiller, J. Crassous, and M. Raynal, 
\textit{Possible chemical and physical scenarios towards biological homochirality}, 
Chem. Soc. Rev. \textbf{51}, 3436 (2022).

\bibitem{Sasselov23}
D. D. Sasselov, J. P. Grotzinger, and J. D. Sutherland, 
\textit{The origin of life as a planetary phenomenon}, 
Sci. Adv. \textbf{6}, eaax341 (2023).

\bibitem{Ozturk23}
S. F. Ozturk, Z. Liu, J. D. Sutherland, and D. D. Sasselov, 
\textit{Origin of biological homochirality by crystallization of
an RNA precursor on a magnetic surface}, 
Sci. Adv. \textbf{9}, eadg8274 (2023).
	
\bibitem{Chang18}
G. Chang, B. J. Wieder, F. Schindler, D. S. Sanchez,
I. Belopolski, S.-M. Huang, 
B. Singh, D. Wu, T.-R. Chang, T. Neupert, S.-Y. Xu, H. Lin, and M Zahid Hasan, 
\textit{Topological quantum properties of chiral crystals}, 
Nat. Mater. \textbf{17}, 978 (2018).

\bibitem{Fecher22}
G.H. Fecher, J. Kübler, and C. Felser, 
\textit{Chirality in the Solid State: Chiral Crystal Structures in Chiral and Achiral Space Groups}, 
Materials \textbf{15}, 5812 (2022).

\bibitem{Felser23}
C. Felser and J. Gooth, 
\textit{Topology and Chirality}, 
Chiral Matter 115 (2023).

\bibitem{Rikken97}
G.L.J.A. Rikken and E. Raupach, 
\textit{Observation of magneto-chiral dichroism,}
Nature \textbf{390}, 493 (1997).

\bibitem{Bordacs12}
S. Bord\'{a}cs, I. K\'{e}zsm\'{a}rki, D. Szaller, L. Demk\'{o}, N. Kida, H. Murakawa, 
Y. Onose, R. Shimano, T. R\~{o}\~{o}m, U. Nagel, S. Miyahara, N. Furukawa, and Y. Tokura, 
\textit{Chirality of matter shows up via spin excitations},
Nat. Phys. \textbf{8}, 734 (2012).

\bibitem{Kezsmarki14}
I. K\'{e}zsm\'{a}rki, D. Szaller, S. Bord\'{a}cs, V. Kocsis, Y. Tokunaga, Y. Taguchi, H. Murakawa, 
Y. Tokura, H. Engelkamp, T. R\~{o}\~{o}m, and U. Nagel, 
\textit{One-way transparency of four-coloured spin-wave excitations in multiferroic materials}, 
Nat. Commun. \textbf{5}, 3203 (2014).

\bibitem{Zhu18}
H. Zhu, J. Yi, M.-Y. Li, J. Xiao, L. Zhang, C.-W. Yang, R. A. Kaindl, 
L.-J. Li, Y. Wang, and
X. Zhang,
\textit{Observation of chiral phonons},
Science \textbf{359}, 6375 (2018).

\bibitem{Nomura19}
T. Nomura, X.-X. Zhang, S. Zherlitsyn, J. Wosnitza, Y. Tokura, N. Nagaosa, and S. Seki, 
\textit{Phonon Magnetochiral Effect}, 
Phys. Rev. Lett. \textbf{122}, 145901 (2019).

\bibitem{Guo22}
C. Guo, C. Putzke, S. Konyzheva, X. Huang, M. Gutierrez-Amigo, I. Errea, D. Chen, 
M. G. Vergniory, C. Felser, M. H. Fischer, T. Neupert, and P. J. W. Moll, 
\textit{Switchable chiral transport in charge-ordered kagome metal CsV$_3$Sb$_5$},
Nature \textbf{611}, 461 (2022).

\bibitem{Yokouchi20}
T. Yokouchi, F. Kagawa, M. Hirschberger, Y. Otani, N. Nagaosa, and Y. Tokura,  
\textit{Emergent electromagnetic induction in a helical-spin magnet},
Nature \textbf{586}, 232 (2020).

\bibitem{Cheong22}
S.-W. Cheong and X. Xu, 
\textit{Magnetic chirality}, 
npj Quant. Mater. \textbf{7}, 40 (2022). 

\bibitem{Muehlbauer09}
S. M\"uhlbauer, B. Binz, F. Jonietz, C. Pfleiderer, A. Rosch, A. Neubauer, 
R. Georgii, and P. B\"oni,  
\textit{Skyrmion Lattice in a Chiral Magnet},
Science \textbf{323}, 915 (2009).

\bibitem{Tokura21}
Y. Tokura and N. Kanazawa, 
\textit{Magnetic Skyrmion Materials}, 
Chem. Rev. \textbf{121}, 2857 (2021).

\bibitem{Kurumaji19}
T. Kurumaji, T. Nakajima, M. Hirschberger, A. Kikkawa, Y. Yamasaki, H. Sagayama, 
H. Nakao, Y. Taguchi, T.-h. Arima and Y. Tokura, 
\textit{Skyrmion lattice with a giant topological Hall effect in a frustrated triangular-lattice magnet}, 
Science \textbf{365}, 914 (2019). 

\bibitem{Hirschberger19}
M. Hirschberger, T. Nakajima, S. Gao, L. Peng, A. Kikkawa, T. Kurumaji, M. Kriener, 
Y. Yamasaki, H. Sagayama, H. Nakao, K. Ohishi, K. Kakurai, Y. Taguchi, 
X. Yu, T.-h. Arima, and Y. Tokura, 
\textit{Skyrmion phase and competing magnetic orders on a breathing kagomé lattice}, 
Nat. Commun. \textbf{10}, 5831 (2019).

\bibitem{Saito08}
M. Saito, K. Ishikawa, K. Taniguchi, and T. Arima, 
\textit{Magnetic Control of Crystal Chirality and the Existence of a 
	Large Magneto-Optical Dichroism Effect in CuB$_2$O$_4$}, 
Phys. Rev. Lett. \textbf{101}, 117402 (2008).

\bibitem{Tokura18}
Y. Tokura and N. Nagaosa, 
\textit{Nonreciprocal responses from non-centrosymmetric quantum materials}, 
Nat. Commun. \textbf{9}, 3740 (2018).

\bibitem{Cheong18}
S.-W. Cheong, D. Talbayev, V. Kiryukhin, and A. Saxena, 
\textit{Broken symmetries, non-reciprocity, and multiferroicity}, 
npj Quant. Mater. \textbf{3}, 19 (2018).

\bibitem{Train08}
C. Train, R. Gheorghe, V. Krstic, L.-M. Chamoreau, N. S. Ovanesyan, 
G. L. J. A. Rikken, M. Gruselle,  and M. Verdaguer, 
\textit{Strong magneto-chiral dichroism in enantiopure chiral ferromagnets}, 
Nat. Mater. \textbf{7}, 729 (2008).

\bibitem{Atzori20}
M. Atzori, G. L. J. A. Rikken, and C. Train, 
\textit{Magneto-Chiral Dichroism: A Playground for Molecular Chemists}, 
Chem. Eur. J. \textbf{26}, 9784 (2020).

\bibitem{Atzori21}
M. Atzori, H. D. Ludowieg, \'A. Valentín-P\'erez, M. Cortijo, I. Breslavetz, 
K. Paillot, P. Rosa, C. Train, J. Autschbach, E. A. Hillard, and G. L. J. A. Rikken, 
\textit{Validation of microscopic magnetochiral
dichroism theory},
Sci. Adv. \textbf{7}, eabg2859 (2021).

\bibitem{Szaller13}
D. Szaller, S. Bord\'{a}cs, and I. K\'{e}zsm\'{a}rki, 
\textit{Symmetry conditions for nonreciprocal light propagation in magnetic crystals},
Phys. Rev. B \textbf{87}, 014421 (2013).

\bibitem{Kezsmarki11}
I. K\'{e}zsm\'{a}rki, N. Kida, H. Murakawa, S. Bord\'{a}cs, Y. Onose, and Y. Tokura, 
\textit{Enhanced Directional Dichroism of Terahertz Light in Resonance with Magnetic 
	Excitations of the Multiferroic Ba$_2$CoGe$_2$O$_7$ Oxide Compound},
Phys. Rev. Lett. \textbf{106}, 057403 (2011). 

\bibitem{Takahashi12}
Y. Takahashi, R. Shimano, Y. Kaneko, H. Murakawa, and Y. Tokura, 
\textit{Magnetoelectric resonance with electromagnons in a perovskite helimagnet}, 
Nat. Phys. \textbf{8}, 121 (2012).

\bibitem{Takahashi13}
Y. Takahashi, Y. Yamasaki, and Y. Tokura, 
\textit{Terahertz Magnetoelectric Resonance Enhanced by Mutual Coupling of Electromagnons}, 
Phys. Rev. Lett. \textbf{111}, 037204 (2013).

\bibitem{Kuzmenko15}
A. M. Kuzmenko, V. Dziom, A. Shuvaev, A. Pimenov, M. Schiebl, A. A. Mukhin, V. Y. Ivanov, 
I. A. Gudim, L. N. Bezmaternykh, and A. Pimenov, 
\textit{Large directional optical anisotropy in multiferroic ferroborate}, 
Phys. Rev. B \textbf{92}, 184409 (2015).

\bibitem{Kezsmarki15}
I. K\'{e}zsm\'{a}rki, U. Nagel, S. Bord\'{a}cs, R. S. Fishman, J. H. Lee, H. T. Yi, 
S.-W. Cheong, and T. R\~{o}\~{o}m, 
\textit{Optical Diode Effect at Spin-Wave Excitations of the Room-Temperature 
	Multiferroic BiFeO$_3$}, 
Phys. Rev. Lett. \textbf{115}, 127203 (2015).

\bibitem{Yu18}
S. Yu, B. Gao, J. W. Kim, S.-W. Cheong, M. K. L. Man, J. Mad\'eo, K. M. Dani, 
and D. Talbayev, 
\textit{High-Temperature Terahertz Optical Diode Effect without Magnetic
Order in Polar FeZnMo$_3$O$_8$},
Phys. Rev. Lett. \textbf{120}, 037601 (2018).

\bibitem{Sirenko21}
A. A. Sirenko, P. Marsik, L. Bugnon, M. Soulier, C. Bernhard, T. N. Stanislavchuk, 
X. Xu, and S.-W. Cheong, 
\textit{Total Angular Momentum Dichroism of the Terahertz Vortex Beams at the 
	Antiferromagnetic Resonances},
Phys. Rev. Lett. \textbf{126}, 157401 (2021).

\bibitem{Yokosuk20}
M. O. Yokosuk, H.-S. Kim, K. D. Hughey, J. Kim, A. V. Stier, K. R. O’Neal, 
J. Yang, S. A. Crooker, K. Haule, S.-W. Cheong, D. Vanderbilt, and J. L. Musfeldt, 
\textit{Nonreciprocal directional dichroism of a chiral magnet in the
visible range}, 
npj Quant. Mater. \textbf{5}, 20 (2020).

\bibitem{Park22}
K. Park, M. O. Yokosuk, M. Goryca, J. J. Yang, S. A. Crooker, S.-W. Cheong, K. Haule, 
D. Vanderbilt, H.-S. Kim, and J. L. Musfeldt, 
\textit{Nonreciprocal directional dichroism at telecom wavelengths}, 
npj Quant. Mat. \textbf{7}, 38 (2022).

\bibitem{Sessoli15}
R. Sessoli, M.-E. Boulon, A. Caneschi, M. Mannini, L. Poggini, F. Wilhelm, 
and A. Rogalev,  
\textit{Strong magneto-chiral dichroism in a paramagnetic molecular
helix observed by hard X-rays}, 
Nat. Phys. \textbf{11}, 69 (2015).

\bibitem{Pop14}
F. Pop, P. Auban-Senzier, E. Canadell, G. L. J. A. Rikken, and N. Avarvari, 
\textit{Electrical magnetochiral anisotropy in a bulk chiral molecular conductor}, 
Nat. Comm. \textbf{5}, 3757 (2014).

\bibitem{Wu22} 
Y. Wu, Q. Wang, X. Zhou, J. Wang, P. Dong, J. He, Y. Ding, B. Teng, Y. Zhang, 
Y. Li, C. Zhao, H. Zhang, J. Liu, Y. Qi, K. Watanabe, T. Taniguchi, and J. Li, 
\textit{Nonreciprocal charge transport in topological kagome superconductor CsV$_3$Sb$_5$}, 
npj Quant. Mater. \textbf{7}, 10 (2022).

\bibitem{Onoda08}
M. Onoda, A. S. Mishchenko, and N. Nagaosa, 
\textit{Left-handed spin wave excitation in ferromagnet}, 
J. Phys. Soc. Jpn. \textbf{77}, 013702 (2008).

\bibitem{Jenni22}
K. Jenni, S. Kunkem\"oller, W. Schmidt, P. Steffens, A. A. Nugroho, and M. Braden, 
\textit{Chirality of magnetic excitations in ferromagnetic SrRuO$_3$}, 
Phys. Rev. B \textbf{105}, L180408 (2022).

\bibitem{Daniels18}
M. W. Daniels, R. Cheng, W. Yu, J. Xiao, and D. Xiao, 
\textit{Nonabelian magnonics in antiferromagnets},
Phys. Rev. B \textbf{98}, 134450 (2018).

\bibitem{Nambu20}
Y. Nambu, J. Barker, Y. Okino, T. Kikkawa, Y. Shiomi, M. Enderle, T. Weber, B. Winn, 
M. Graves-Brook, J. M. Tranquada, T. Ziman, M. Fujita, G. E. W. Bauer, E. Saitoh, 
and K. Kakurai, 
\textit{Observation of Magnon Polarization},
Phys. Rev. Lett. \textbf{125}, 027201 (2020). 

\bibitem{Liu22}
Y. Liu, Z. Xu, L. Liu, K. Zhang, Y. Meng, Y. Sun, P. Gao, H.-W. Zhao, Q. Niu, and J. Li,  
\textit{Switching magnon chirality in artificial ferrimagnet},
Nat. Commun. \textbf{13}, 1264 (2022).

\bibitem{Sahasrabudhe23}
A. Sahasrabudhe, M. A. Prosnikov, T. C. Koethe, P. Stein, V. Tsurkan, 
A. Loidl, M. Gr\"uninger, H. Hedayat, and P. H. M. van Loosdrecht, 
\textit{Chiral excitations and the intermediate-field spin-liquid regime 
	in the Kitaev magnet $\alpha$-RuCl$_3$}, 
arxiv:2305.03400.

\bibitem{Ishito23}
K. Ishito, H. Mao, Y. Kousaka, Y. Togawa, S. Iwasaki, T. Zhang, 
S. Murakami, J.-i. Kishine, and T. Satoh, 
\textit{Truly chiral phonons in $\alpha$-HgS}, 
Nat. Phys. \textbf{19}, 35 (2023).

\bibitem{Glazer06}
A.M. Glazer and K.G. Cox, 
\textit{International Table for Crystallography}, 
Vol.\ D, chap.\ 1.6, pp. 150-177 (2006).

\bibitem{Yokosuk16}
M. O. Yokosuk, A. al-Wahish, S. Artyukhin, K. R. O’Neal, D. Mazumdar, P. Chen, J. Yang, 
Y. S. Oh, S. A. McGill, K. Haule, S.-W. Cheong, D. Vanderbilt, and J. L. Musfeldt, 
\textit{Magnetoelectric Coupling through the Spin Flop Transition in Ni$_3$TeO$_6$}, 
Phys. Rev. Lett. \textbf{117}, 147402 (2016).

\bibitem{Oh14}
Y.S. Oh, S. Artyukhin, J.J. Yang, V. Zapf, J.W. Kim, D. Vanderbilt, and S.-W. Cheong, 
\textit{Non-hysteretic colossal magnetoelectricity in a collinear antiferromagnet}, 
Nat. Commun. \textbf{5}, 3201 (2014).

\bibitem{Zivkovic10}
I. \v{Z}ivkovi\'{c}, K. Pr\v{s}a, O. Zaharko, and H. Berger, 
\textit{Ni$_3$TeO$_6$ - a collinear antiferromagnet with ferromagnetic honey\-comb planes}, 
J. Phys.: Condens. Matter \textbf{22}, 056002 (2010).

\bibitem{Skiadopoulou17}
S. Skiadopoulou, F. Borodavka, C. Kadlec, F. Kadlec, M. Retuerto, Z. Deng, M. Greenblatt, 
and S. Kamba, 
\textit{Magnetoelectric excitations in multiferroic Ni$_3$TeO$_6$}, 
Phys. Rev. B \textbf{95}, 184435 (2017).

\bibitem{Becker06}
R. Becker and H. Berger, 
\textit{Reinvestigation of Ni$_3$TeO$_6$}, 
Acta Cryst. \textbf{62}, i222 (2006).

\bibitem{Peedu22}
L. Peedu, V. Kocsis, D. Szaller, B. Forrai, S. Bord\'acs, I. K\'ezsm\'arki, 
J. Viirok, U. Nagel, B. Bern\'ath, D. L. Kamenskyi, A. Miyata, O. Portugall, 
Y. Tokunaga, Y. Tokura, Y. Taguchi, and T. R\~o\~om, 
\textit{Terahertz spectroscopy of spin excitations in magnetoelectric 
	LiFePO$_4$ in high magnetic fields}, 
Phys. Rev. B \textbf{106}, 134413 (2022).

\bibitem{Peedu19}
L. Peedu, V. Kocsis, D. Szaller, J. Viirok, U. Nagel, T. R\~o\~om, D. G. Farkas, 
S. Bord\'acs, D. L. Kamenskyi, U. Zeitler, Y. Tokunaga, Y. Taguchi, 
Y. Tokura, and I. K\'ezsm\'arki, 
\textit{Spin excitations of magnetoelectric LiNiPO$_4$ in multiple magnetic phases}, 
Phys. Rev. B \textbf{100}, 024406 (2019).

\bibitem{Kocsis18}
V. Kocsis, K. Penc, T. R\~o\~om, U. Nagel, J. Vít, J. Romh\'anyi, Y. Tokunaga, 
Y. Taguchi, Y. Tokura, I. K\'ezsm\'arki, and S. Bord\'acs, 
\textit{Identification of Antiferromagnetic Domains Via the Optical Magnetoelectric Effect}, 
Phys. Rev. Lett. \textbf{121}, 057601 (2018).

\bibitem{Szaller17}
D. Szaller, V. Kocsis, S. Bord\'acs, T. Feh\'er, T. R\~o\~om, U. Nagel, 
H. Engelkamp, K. Ohgushi, and I. K\'ezsm\'arki, 
\textit{Magnetic resonances of multiferroic TbFe$_3$(BO$_3$)$_4$},
Phys. Rev. B \textbf{95}, 024427 (2017).

\bibitem{Pimenov06}
A. Pimenov, A. A. Mukhin, V. Yu. Ivanov, V. D. Travkin, A. M. Balbashov, 
and A. Loidl, 
\textit{Possible evidence for electromagnons in
multiferroic manganites}, 
Nat. Phys. \textbf{2}, 97 (2006).

\bibitem{Valdes09}
R. Vald\'es Aguilar, M. Mostovoy, A. B. Sushkov, C. L. Zhang, Y. J. Choi, 
S-W. Cheong, and H. D. Drew, 
\textit{Origin of Electromagnon Excitations in Multiferroic $R$MnO$_3$}, 
Phys. Rev. Lett. \textbf{102}, 047203 (2009).

\bibitem{Finger14}
T. Finger, K. Binder, Y. Sidis, A. Maljuk, D. N. Argyriou, and M. Braden, 
\textit{Magnetic order and electromagnon excitations in DyMnO$_3$ studied
by neutron scattering experiments}, 
Phys. Rev. B \textbf{90}, 224418 (2014).

\bibitem{Tokura14}
Y. Tokura, S. Seki, and N. Nagaosa, 
\textit{Multiferroics of spin origin}, 
Rep. Prog. Phys. \textbf{77}, 076501 (2014).

\bibitem{Reschke22}
S. Reschke, D. G. Farkas, A. Strinić, S. Ghara, K. Guratinder, O. Zaharko, 
L. Prodan, V. Tsurkan, D. Szaller, S. Bordács, J. Deisenhofer, and I. K\'ezsm\'arki,  
\textit{Confirming the trilinear form of the optical magnetoelectric effect 
	in the polar honeycomb antiferromagnet Co$_2$Mo$_3$O$_8$},
npj Quant. Mater. \textbf{7}, 1 (2022).

\bibitem{Csizi20}
B. Csizi, S. Reschke, A. Strini\'c, L. Prodan, V. Tsurkan, I. K\'ezsm\'arki, 
and J. Deisenhofer, 
\textit{Magnetic and vibronic terahertz excitations in Zn-doped Fe$_2$Mo$_3$O$_8$},
Phys. Rev. B \textbf{102}, 174407 (2020).

\bibitem{Bayrakci13}
S. P. Bayrakci, D. A. Tennant, Ph. Leininger, T. Keller, M. C. R. Gibson, 
S. D. Wilson, R. J. Birgeneau, and B. Keimer, 
\textit{Lifetimes of Antiferromagnetic Magnons in Two and Three Dimensions:
Experiment, Theory, and Numerics}, 
Phys. Rev. Lett. \textbf{111}, 017204 (2013).

\bibitem{Weymann21}
L. Weymann, A. Shuvaev, A. Pimenov, A. A. Mukhin, and D. Szaller, 
\textit{Magnetic equivalent of electric superradiance in yttrium-iron-garnet films}, 
Commun. Phys. \textbf{4}, 97 (2021).

\bibitem{Wang15}
X. Wang, F.-T. Huang, J. Yang, , Y.S. Oh, and S.-W. Cheong, 
\textit{Interlocked chiral/polar domain walls and large optical rotation in Ni$_3$TeO$_6$},
APL Mater. \textbf{3}, 076105 (2015).

\bibitem{Zhang09}
S. Zhang, Y.-S. Park, J. Li, X. Lu, W. Zhang, and X. Zhang, 
\textit{Negative Refractive Index in Chiral Metamaterials}, 
Phys. Rev. Lett. \textbf{102}, 023901 (2009). 

\bibitem{Smith76}
D. Y. Smith, 
\textit{Superconvergence and sum rules for the optical constants: 
	Natural and magneto-optical activity}, 
Phys. Rev. B \textbf{13}, 5303 (1976).

\bibitem{Zupan71}
J. Zupan, D. Kolar, and V. Urbanc, 
\textit{Magnetic properties of Ni$_3$TeO$_6$}, 
Mat. Res. Bull. \textbf{6}, 1353 (1971).

\bibitem{Kuzmenko14}
A. M. Kuzmenko, A. Shuvaev, V. Dziom, A. Pimenov, M.Schiebl, A. A. Mukhin, V. Yu. Ivanov, 
L. N. Bezmaternykh,and A. Pimenov, 
\textit{Giant gigahertz optical activity in multiferroic ferroborate}, 
Phys. Rev. B \textbf{89}, 174407 (2014).

\bibitem{Kuzmenko19}
A. M. Kuzmenko, V. Dziom, A. Shuvaev, A. Pimenov, D. Szaller, A. A. Mukhin, V. Yu. Ivanov, and A. Pimenov, 
\textit{Sign change of polarization rotation under time or space inversion in magnetoelectric YbAl$_3$(BO$_3$)$_4$}, 
Phys. Rev. B \textbf{99}, 224417 (2019).

\bibitem{Iguchi21}
S. Iguchi, R. Masuda, S. Seki, Y. Tokura, and Y. Takahashi, 
\textit{Enhanced gyrotropic birefringence and natural optical activity 
	on electromagnon resonance in a helimagnet},
Nat. Commun. \textbf{12}, 6674 (2021).

\bibitem{Sankar13}
R. Sankar, G. J. Shu, B. Karunakara Moorthy, R. Jayaveld, F. C. Chou,
\textit{Growing of fixed orientation plane of single crystal using the flux growth technique 
	and ferrimagnetic ordering in Ni$_3$TeO$_6$ of stacked 2D honeycomb rings},
Dalton Trans. \textbf{42}, 10439 (2013).

\bibitem{Kim15}
J. W. Kim, S. Artyukhin, E. D. Mun, M. Jaime, N. Harrison, A. Hansen, 
J. J. Yang, Y. S. Oh, D. Vanderbilt, V. S. Zapf, and S.-W. Cheong, 
\textit{Successive Magnetic-Field-Induced Transitions and Colossal Magnetoelectric
	Effect in Ni$_3$TeO$_6$},
Phys. Rev. Lett. \textbf{115}, 137201 (2015).

\bibitem{Maluski22}
D. Maluski, I. C\'amara Mayorga, J. Hemberger, and M. Gr\"{u}ninger, 
\textit{Terahertz Measurements on Subwavelength-Size Samples Down to the Tunneling Limit}, 
J. Infrared Millim. Terahertz Waves \textbf{43}, 314 (2022).

\bibitem{Roggenbuck10}
A. Roggenbuck, H. Schmitz, A. Deninger, I. C\'amara Mayorga, J. Hemberger, R. G\"usten, 
and M. Gr\"uninger, 
\textit{Coherent broadband continuous-wave terahertz spectroscopy on solid-state samples}, 
New J. Phys. \textbf{12}, 043017 (2010).

\bibitem{Roggenbuck12}
A. Roggenbuck, K. Thirunavukkuarasu, H. Schmitz, J. Marx, A. Deninger, I. C\'amara Mayorga, 
R. G\"usten, J. Hemberger, and M. Gr\"uninger, 
\textit{Using a fiber stretcher as a fast phase modulator in a continuous wave terahertz spectrometer}, 
J. Opt. Soc. Am. B \textbf{29}, 614 (2012).

\bibitem{Roggenbuck13}
A. Roggenbuck, M. Langenbach, K. Thirunavukkuarasu, H. Schmitz, A. Deninger, I. C\'amara Mayorga, 
R. G\"usten, J. Hemberger, and M. Gr\"uninger, 
\textit{Enhancing the stability of a continuous-wave terahertz system by photocurrent normalization}, 
J. Opt. Soc. Am. B \textbf{30}, 1397 (2013).	

\bibitem{Camara06}
I. C\'amara Mayorga, P. Mu\~{n}oz Pradas, E. A. Michael, M. Mikulics, A. Schmitz, 
P. van der Wal, C. Kaseman, R. G\"usten, K. Jacobs, M. Marso, H. L\"uth, and P. Kordo\v{s}, 
\textit{Terahertz photonic mixers as local oscillators for hot electron bolometer
	and superconductor-insulator-superconductor astronomical receivers}, 
J. Appl. Phys. \textbf{100}, 043116 (2006).

\bibitem{Furuya09}
T. Furuya, K. Maeda, K. Yamamoto, T. Nakashima, T. Inoue, M. Hangyo, and M. Tani, 
\textit{Broadband polarization properties of photoconductive spiral antenna}, 
34$^{\rm th}$ Int. Conf. Infrared, Millimeter, and Terahertz Waves, Busan, Korea (South), 
pp. 1-2 (2009), DOI: 10.1109/ICIMW.2009.5324916.

\bibitem{Garufo16}
A. Garufo, N. Llombart and A. Neto, 
\textit{Radiation of Logarithmic Spiral Antennas in the Presence of Dense Dielectric Lenses}, 
IEEE Trans. Antennas Propag. \textbf{64}, 4168 (2016). 


\end{thebibliography}
\end{document}